\documentclass[aps,prd,twocolumn,superscriptaddress,nofootinbib]{revtex4-2}

\usepackage{amsmath,amssymb,amsthm,bm,graphicx,xcolor,subfigure}
\usepackage[colorlinks=true,linkcolor=blue,citecolor=blue,urlcolor=blue]{hyperref}

\begin{document}

\title{Spontaneous Deformation of an AdS Spherical Black Hole}

\author{Zhuan Ning}
\email{ningzhuan17@mails.ucas.ac.cn}
\affiliation{School of Physical Sciences, University of Chinese Academy of Sciences, Beijing 100049, China}

\author{Qian Chen}
\email{chenqian192@mails.ucas.ac.cn}
\thanks{Corresponding author}
\affiliation{School of Physical Sciences, University of Chinese Academy of Sciences, Beijing 100049, China}

\author{Yu Tian}
\email{ytian@ucas.ac.cn}
\thanks{Corresponding author}
\affiliation{School of Physical Sciences, University of Chinese Academy of Sciences, Beijing 100049, China}
\affiliation{Institute of Theoretical Physics, Chinese Academy of Sciences, Beijing 100190, China}

\author{Xiaoning Wu}
\email{wuxn@amss.ac.cn}
\thanks{Corresponding author}
\affiliation{Institute of Mathematics, Chinese Academy of Sciences, Beijing 100190, China}

\author{Hongbao Zhang}
\email{hongbaozhang@bnu.edu.cn}
\thanks{Correponding author}
\affiliation{Department of Physics, Beijing Normal University, Beijing 100875, China}

\begin{abstract}
In this study, we investigate the real-time dynamics during the spontaneous deformation of an unstable spherical black hole in asymptotically anti-de Sitter (AdS) spacetime. For the initial value, the static solutions with spherical symmetry are obtained numerically, revealing the presence of a spinodal region in the phase diagram. From the linear stability analysis, we find that only the central part of such a thermodynamically unstable spinodal region leads to the emergence of a type of axial instability. To trigger the dynamical instability, an axial perturbation is imposed on the scalar field. As a result, by the fully nonlinear dynamical simulation, the spherical symmetry of the gravitational system is broken spontaneously, leading to the formation of an axisymmetric black hole.
\end{abstract}

\maketitle

\section{Introduction}

In the framework of general relativity, a four-dimensional, static, vacuum, asymptotically flat black hole can be fully characterized by a unique parameter known as the Arnowitt-Deser-Misner mass \cite{Chrusciel:2012jk}. 
Furthermore, the topology of the event horizon of such a black hole must be a two-dimensional sphere $S^2$ \cite{Hawking:1971vc,Hawking:1973uf,Friedman:1993ty,Chrusciel:1994tr}. However, some solutions with additional conserved quantities and other horizon topologies will survive if one takes into account some additional matter fields, higher dimensions, or different spacetime asymptotics.

In particular, in the asymptotically AdS spacetime, the horizon topology of black holes is not limited to $S^2$. Solutions that asymptotically approach a local AdS spacetime can possess a horizon with planar or hyperbolic topology. The properties of these ``black holes" have been studied extensively \cite{Vanzo:1997gw,Birmingham:1998nr,Aros:2000ij}.
In particular, in the context of the AdS/CFT correspondence \cite{Maldacena:1997re,Gubser:1998bc,Witten:1998qj,Witten:1998zw}, the planar black brane is the main object of interest \cite{Chesler:2010bi}, which is naturally dual to a strongly coupled conformal field theory in the Minkowski spacetime with one less dimension. Interestingly, an inhomogeneous phase-separated black brane can dynamically arise from an initially homogeneous black brane in certain gravitational models \cite{Attems:2017ezz,Janik:2017ykj,Attems:2019yqn,Bellantuono:2019wbn,Chen:2022cwi}, corresponding holographically to a gauge theory with a first-order thermal phase transition.
The thermodynamic relations of the equilibrium states \cite{Gubser:2008ny,Dias:2017uyv}, the linear stability analysis \cite{Janik:2015iry,Gursoy:2016ggq}, and the nonlinear dynamics \cite{Gursoy:2016ggq,Attems:2017ezz,Janik:2017ykj,Attems:2019yqn,Bellantuono:2019wbn,Chen:2022cwi,Chen:2022tfy} have been investigated in these models. Specifically, the initial states are afflicted by the so-called spinodal instability, and the time evolution of this kind of instability can be well approximated by second-order hydrodynamics \cite{Attems:2017ezz,Attems:2019yqn}.
On the gravity side, the spinodal instability is a long wavelength instability, similar to the Gregory-Laflamme instability \cite{Gregory:1993vy,Gregory:1994bj}, though there are also significant distinctions between them \cite{Attems:2017ezz}. In addition, these gravitational models allow a rich set of static lumpy black brane solutions when the size of the box is varied \cite{Bea:2020ees}.
These configurations are not triggered by an explicit inhomogeneous external source. Instead, they are formed by spontaneously breaking the translational symmetry.

However, the story may be more intriguing for solutions with a topologically spherical horizon that asymptotically approach a global AdS spacetime. 
By including an azimuthal winding number in the matter field ansatz, the static black holes with only axial symmetry were constructed in Einstein-scalar field and Einstein-Yang-Mills theories \cite{Kichakova:2015nni}.
By turning on a spatially dependent external source, static axisymmetric black holes within a dipole soliton were constructed in Einstein-Maxwell theory \cite{Costa:2015gol}.
Later on, this construction was generalized to higher multipoles, whereby the static black holes without any continuous spatial symmetry were obtained \cite{Herdeiro:2016plq,Herdeiro:2020saw}.

On the other hand, inspired by the aforementioned holographic model for the first-order phase transition, we are tempted to suspect that there may exist static black hole configurations with only axial symmetry in the absence of a winding number and a spatially dependent external source. Put it another way, there may exist some spherically symmetric black holes, which are unstable and will dynamically evolve to a stable configuration with only axial symmetry remaining.
This kind of process is
similar to the case of the higher dimensional charged black holes in the de Sitter spacetime \cite{Konoplya:2008au},
where the perturbed black holes will undergo a deformation, transforming into another solution.

The main purpose of this work is to study the spontaneous breaking of spherical symmetry of AdS black holes with spherical horizons in a certain gravitational model. As alluded to above, it is expected that these black holes have a lot in common with their planar counterparts -- those in a holographic model with a first-order phase transition. Particularly, some of them suffer from spinodal instability and will undergo a spontaneous deformation process after certain perturbations, leading to the formation of black holes with only axial symmetry. On the other hand, spherical black holes also demonstrate some new features, which will be documented in our paper.

The rest of the paper is structured as follows. In Sec. \ref{sec: model} we introduce our gravitational model, and reveal the phase diagram structures for static solutions with spherical symmetry. In Sec. \ref{sec: linear} we perform the linear stability analysis of the equilibrium states located in the spinodal region, revealing the existence of a type of axial instability. In Sec. \ref{sec: nonlinear} we trigger the instability by adding an axisymmetric scalar perturbation and conduct the nonlinear time evolution to simulate the dynamical process till the formation of the final stable configuration. Finally, we conclude our paper with some discussions in Sec. \ref{sec: conclusion}. In the appendices, we present the general formalism of numerical relativity under the Bondi-Sachs-like metric and further details of the evolution scheme we have used in this paper.

\section{Gravity model and static solutions}\label{sec: model}

\subsection{Gravity model and metric ansatz}

We consider the Einstein gravity coupled to a real scalar field with a self-interacting potential in the four-dimensional asymptotically AdS spacetime, described by the Lagrangian density
\begin{equation}\label{eq: Lagrangian density}
    \mathcal{L}=R-\frac{1}{2}\nabla_\mu\phi\nabla^\mu\phi-V(\phi).
\end{equation}
For simplicity, we set the AdS radius $L$ to the unit and focus on a scalar field with the mass squared $m^2=-2$ within the Breitenlohner-Freedman bound \cite{Breitenlohner:1982jf}. In order to obtain unstable spherically symmetric black holes, the scalar potential is specified as
\begin{equation}
    V(\phi)=-6\cosh\left(\frac{\phi}{\sqrt{3}}\right)-\frac{\phi^4}{5}.
\end{equation}
This form of potential is the same as in \cite{Janik:2017ykj}, though other forms of $V(\phi)$ can also lead to qualitatively similar results. The field equations to be solved can be extracted from the variation of the Lagrangian density (\ref{eq: Lagrangian density}), as follows
\begin{equation}\label{eq: EOM}
    \begin{aligned}
        R_{\mu\nu}-\frac{1}{2}Rg_{\mu\nu}&=\frac{1}{2}\nabla_\mu\phi\nabla_\nu\phi-\left(\frac{1}{4}(\nabla\phi)^2+\frac{1}{2}V(\phi)\right)g_{\mu\nu},\\
        \nabla^\mu\nabla_\mu\phi&=\frac{dV(\phi)}{d\phi}.
    \end{aligned}
\end{equation}

To perform a dynamical simulation of the spontaneous deformation of a spherically symmetric black hole, we adopt the ingoing Bondi-Sachs-like coordinates \cite{Bondi:1962px,Sachs:1962wk,Balasubramanian:2013yqa} as the metric ansatz\footnote{See also, e.g. \cite{Cao:2013ema,He:2015wfa} and references therein for discussions about their outgoing counterparts.}
\begin{widetext}
    \begin{equation}\label{eq: metric}
        ds^2=\frac{L^2}{z^2}(-[fe^{-\chi}-e^{A}\xi^2]dv^2-2e^{-\chi}dvdz-2\xi e^Advd\theta+e^Ad\theta^2+e^{-A}\sin^2\theta d\varphi^2),
    \end{equation}
\end{widetext}
where $L$ has been fixed to the unit and the compactification coordinate $z=r^{-1}$ is introduced to constrain the computational domain to be finite. 
For simplicity, we preserve the axial symmetry in the $\varphi$ direction such that all metric components and the scalar field are functions of $(v,z,\theta)$.
With such an ansatz, the metric components and the scalar field have the following asymptotic behavior near the AdS boundary ($z=0$):
\begin{align}
    \chi&=\frac{\phi_1^2}{8}z^2+O(z^3),\\
    \xi&=\xi_3 z^3+O(z^4),\\
    f&=1+\left(\frac{\phi_1^2}{8}+1\right)z^2+f_3 z^3+O(z^4),\\
    A&=O(z^3),\\
    \phi&=\phi_1 z+\phi_2 z^2+O(z^3).
\end{align}
Note that we have chosen the gauge $\chi|_{z=0}=0$, $\xi|_{z=0}=0$, and $A|_{z=0}=0$ in this paper. Here, the scalar source $\phi_1$ is a boundary freedom, and the response $\phi_2$ can only be determined after solving the full equations of motion (EOM). Different from the case of planar topology where the scalar source is only a scaling freedom, here the different values of the source will result in the physical scenarios with substantial distinctions, 
which will be seen later.

\subsection{Numerical procedure}

For the static solutions with spherical symmetry, we find that the fields $\xi$, $A$ can be turned off, and the remaining fields $\chi$, $f$, $\phi$ are functions of $z$ only, which can be solved efficiently by the Newton-Raphson iteration algorithm with appropriate boundary conditions. 
Consequently, the EOM (\ref{eq: EOM}) degenerate to
\begin{align}
    \label{eq: chi_static} \chi'&=\frac{z}{4}\phi'^2,\\
    \label{eq: f_static} \left(\frac{f}{z^{3}}\right)^{\prime}&=\frac{L^{2}}{2z^{4}}e^{-\chi}V(\phi)-\frac{1}{z^{2}}e^{-\chi},\\
    \label{eq: phi_static} \frac{z^2}{2}\left(\frac{f\phi'}{z^2}\right)'&=\frac{L^2}{2z^2}e^{-\chi}\frac{dV(\phi)}{d\phi},\\
    \label{eq: error_static} \frac{(f\chi')'}{2}&=\frac{f \phi'^{2}}{4}+\frac{3f}{z^{2}}-\frac{2f'}{z}+\frac{f''}{2}+\frac{L^{2}}{2z^{2}}e^{-\chi}V(\phi).
\end{align}

To obtain a static, spherically symmetric black hole solution with an event horizon at $z=z_h$, we take the computational domain $[0,z_0]$ of coordinate $z$, with $z_0$ chosen to be slightly larger than $z_h$.
A solution to the system of equations (\ref{eq: chi_static})-(\ref{eq: error_static}) cannot be determined until the boundary conditions are also specified.
For equation (\ref{eq: chi_static}), we impose the boundary condition $\chi|_{z=0}=0$, which is just a gauge choice.
For equation (\ref{eq: f_static}), the boundary condition can be imposed either at the AdS boundary or at the inner boundary. Since the location of the event horizon $z_h$ of a static, spherically symmetric black hole is determined by the condition $f(z_h)=0$ in the metric ansatz (\ref{eq: metric}), we choose the inner boundary condition $f|_{z=z_0}=f_0$, where $f_0$ is a negative constant given by hand.
For equation (\ref{eq: phi_static}), we impose the boundary condition $\phi'|_{z=0}=\phi_1$, where $\phi_1$ is the scalar source.
The redundant equation (\ref{eq: error_static}) is used to detect numerical errors.

We solve our boundary value problem using a Newton-Raphson algorithm, along with a Chebyshev pseudospectral discretization in the $z$ direction. These methods are reviewed and detailed in the review \cite{Dias:2015nua} and are used, for example, in \cite{Dias:2015pda,Markeviciute:2016ivy,Bea:2020ees}.
Specifically, the above equation set (\ref{eq: chi_static})-(\ref{eq: phi_static}) can be denoted by $\bm{E}[z,\bm{F}]=0$ with $\bm{F}=(\chi,f,\phi)$. The new value $\bm{F}_{i+1}$ is obtained from its value in the previous step $\bm{F}_i$:
\begin{equation}
    \bm{F}_{i+1}=\bm{F}_i-\bm{J}^{-1}(\bm{F}_i)\bm{E}(\bm{F}_i),
\end{equation}
where $\bm{J}=\frac{\delta\bm{E}}{\delta\bm{F}}$ is the functional Jacobian. Once an initial value of $\bm{F}$ is given, we iterate the procedure until the difference $\bm{F}_N-\bm{F}_{N-1}$ is small enough, and consider $\bm{F}_N$ to be a static solution.
Our criterion for convergence is $\max|\bm{F}_{i+1}-\bm{F}_i|<10^{-10}$.
The resulting static solutions have numerical errors of less than $10^{-5}$ outside the horizon, although the errors are larger inside the horizon.

\subsection{Phase diagrams}

After obtaining the spacetime geometry, we want to find the radial position $z_h$ of the event horizon. We first use the barycentric interpolation \cite{trefethen2000spectral} to interpolate the function $f(z)$ to a continuous function, and then find its root using the function \texttt{scipy.optimize.root} in Python.
Once the horizon has been determined, one can easily extract the Hawking temperature of the black hole:
\begin{equation}
    T=\frac{|f'(z_h)|}{4\pi}.
\end{equation}
We use the holographic renormalization procedure \cite{Bianchi:2001kw,Skenderis:2002wp,Elvang:2016tzz} to find the energy-momentum-stress tensor, and its component $-T_v^v$ represents the energy density of the boundary system, which is expressed as follows:
\begin{equation}
    \varepsilon=-f_3+\frac{\phi_{1}\phi_{2}}{6},
\end{equation}
where $f_{3}$ is the coefficient of the cubic term in the asymptotic expansion of the field $f$ near the boundary.

\begin{figure*}
    \subfigure[]{\includegraphics[width=0.45\linewidth]{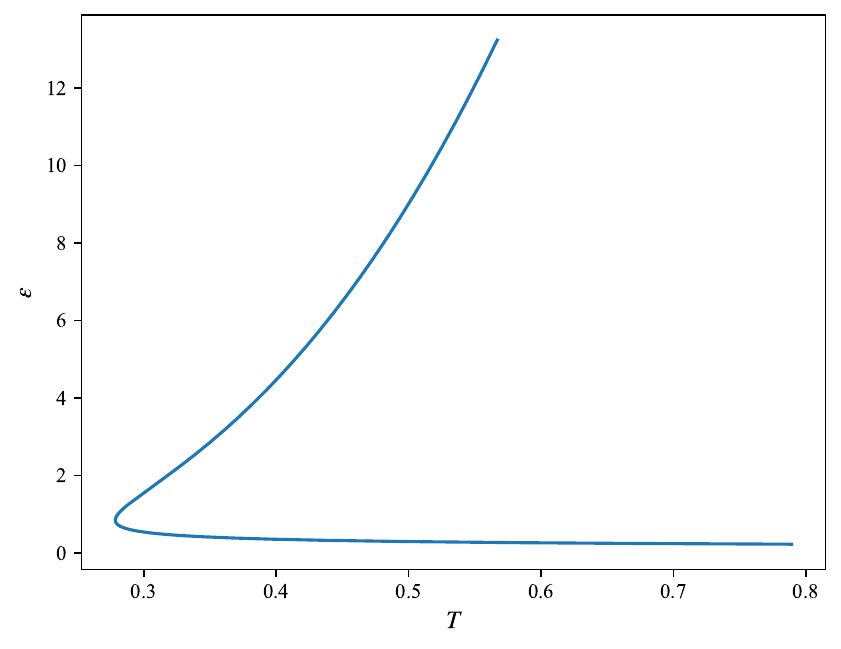}}
    \subfigure[]{\includegraphics[width=0.45\linewidth]{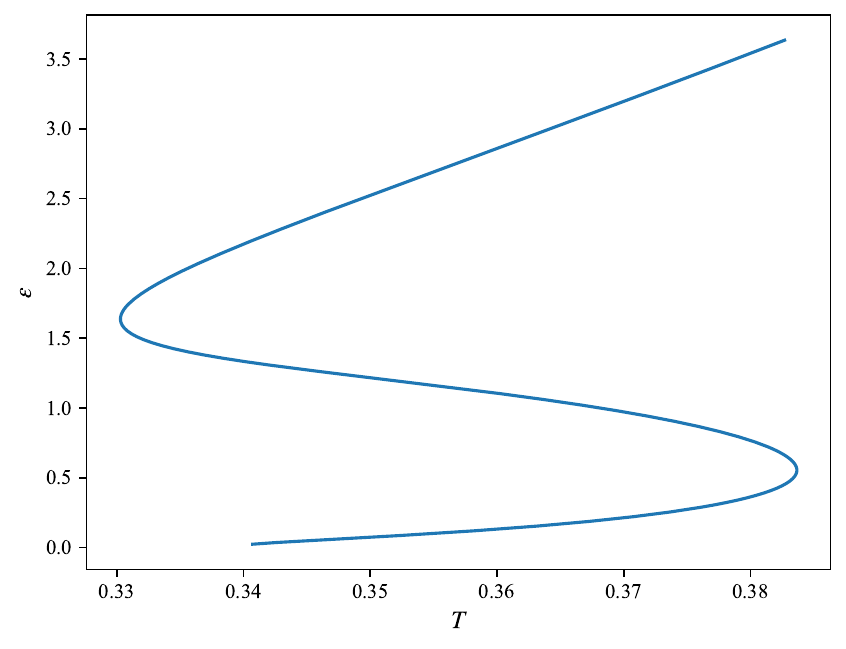}}
    \subfigure[]{\includegraphics[width=0.45\linewidth]{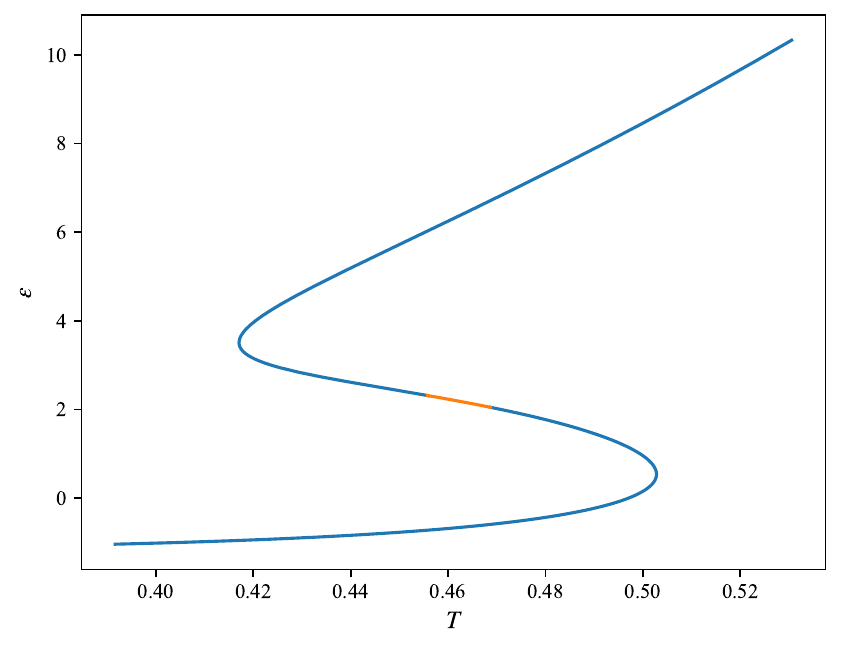}}
    \subfigure[]{\includegraphics[width=0.45\linewidth]{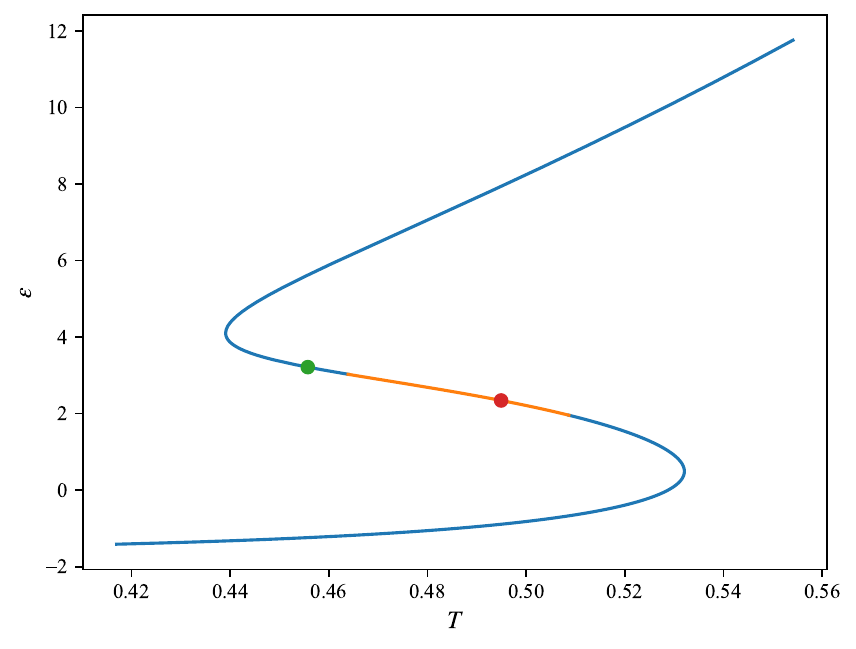}}
    \subfigure[]{\includegraphics[width=0.45\linewidth]{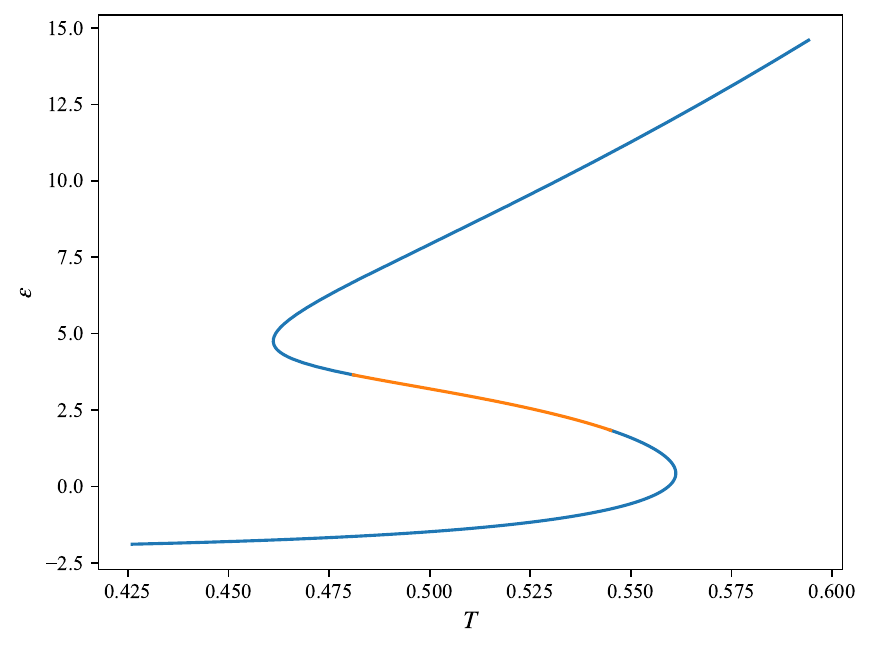}}
    \subfigure[]{\includegraphics[width=0.45\linewidth]{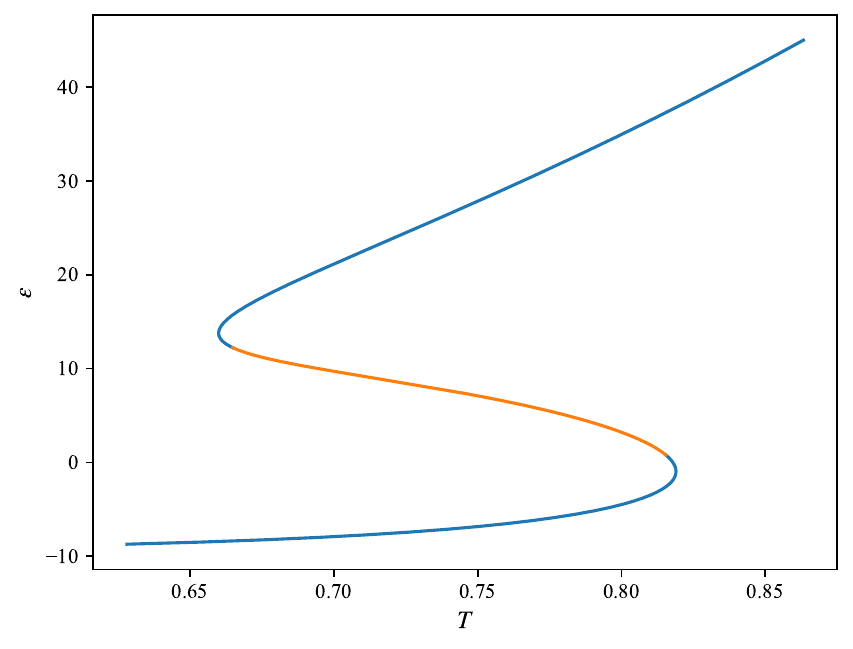}}
    \caption{The temperature dependence of the energy density {for static solutions with scalar source $\phi_1=1$ (a), $1.5$ (b), $1.9$ (c), $2$ (d), $2.1$ (e), and $3$ (f)}. The orange region indicates the states with dynamical instability. The green and red dots represent the chosen initial configurations for the nonlinear dynamical simulation.}
    \label{fig: phase diagram}
\end{figure*}

We take a given value of the scalar source $\phi_1$ and vary $f_0$ within a certain range to obtain a set of static solutions.
The resulting thermodynamic relations of these static solutions for different values of the source $\phi_1=1$, $1.5$, $1.9$, $2$, $2.1$, and $3$ are displayed in Fig. \ref{fig: phase diagram} as the temperature dependence of the energy density.
For sufficiently large sources, such as those equal to or larger than $1.5$, the phase diagrams are qualitatively similar to the case of planar topology \cite{Janik:2017ykj}.
One can observe that these phase diagrams are divided into three branches by two turning points. The branches between the two turning points are the so-called spinodal regions. For a source that is too small, such as $\phi_1=1$, there is only one turning point and two branches of solutions, but we still call the branch below the turning point the spinodal region. The equilibrium states located in these spinodal regions are thermodynamically unstable due to the negative specific heat $c_v=d\varepsilon/dT$.

In the case of planar topology, due to the existence of hydrodynamic modes, such thermodynamic instability results in a class of long-wavelength dynamical instability \cite{Attems:2017ezz,Attems:2019yqn}, similar to the Gregory-Laflamme instability. Meanwhile, states in thermodynamically stable regions are also dynamically stable at the linear level. The consistency between thermodynamic stability and dynamical stability suggests that the Gubser-Mitra conjecture \cite{Gubser:2000ec,Gubser:2000mm} holds for the case of planar topology. In the spherical case, however, even though a thermodynamically unstable region exists, without linear perturbation analysis we do not know whether there are dynamically unstable states.

\section{Linear stability analysis}\label{sec: linear}

\subsection{Numerical procedure}

To reveal the dynamical instability of the thermal phases at the linear level, we compute the quasi-normal modes with an angular quantum number of the gravitational system. The quasi-normal modes are the solutions to the linearized EOM around the background solution, which are characterized by complex frequencies $\omega=\omega_R+i\omega_I$. The dynamical stability of the background solution depends on the imaginary part of the quasi-normal frequencies. If there exists a quasi-normal mode with a positive imaginary part, the perturbations will grow exponentially and push the gravitational system out of equilibrium, otherwise, the perturbations will decay with time.

We use the generalized eigenvalue method to numerically calculate the quasi-normal modes in our system, which was first applied to general relativity in \cite{Dias:2009iu,Dias:2010eu,Monteiro:2009ke}, and has been used extensively in the probe limit \cite{Du:2015zcb,Guo:2018mip} and a wide class of gravitational systems \cite{Destounis:2020pjk,Jansen:2020hfd,Chen:2023eru}. The method is applicable even though the perturbation equations are coupled and the background is a numerical solution. A detailed description and an available Mathematica package of this method can be found in \cite{Jansen:2017oag}. We consider perturbations on the static, spherically symmetric background in the following form
\begin{equation}
    \begin{aligned}
        g_{\mu\nu}(v,z,\theta)&=g_{\mu\nu}^{(0)}(z)+\delta g_{\mu\nu}(v,z,\theta),\\
        \phi(v,z,\theta)&=\phi^{(0)}(z)+\delta \phi(v,z,\theta).
    \end{aligned}
\end{equation}
By inserting this form into the nonlinear EOM (\ref{eq: chi_constraint})-(\ref{eq: f_evolution}) and retaining only the linear terms, we obtain linear perturbation equations for $\delta\chi$, $\delta f$, $\delta\xi$, $\delta A$, and $\delta\phi$. To separate the angular dependence in the perturbation equations, we introduce two new variables
\begin{equation}
    \begin{aligned}
        \delta\Xi &= \frac{1}{\sin\theta}\partial_\theta(\sin\theta \delta\xi)\\
        &= \partial_\theta\delta\xi + \cot\theta\delta\xi,\\
        \delta a &= \frac{1}{\sin\theta}\partial_\theta(\sin\theta (\partial_\theta\delta A+2\cot\theta \delta A))\\
        &= \partial_\theta^2\delta A + 3 \cot\theta\partial_\theta\delta A - 2 \delta A.
    \end{aligned}
\end{equation}
By this substitution, all $\theta$-dependence in the perturbation equations can be transformed into two-dimensional Laplacian operators without $\varphi$-dependence: $\Delta_2=\frac{1}{\sin\theta}\frac{\partial}{\partial\theta}\left(\sin\theta\frac{\partial}{\partial\theta}\right)$, and the perturbation equations are simplified to
\begin{widetext}
    \begin{align}
        \delta\chi'&=\frac{z}{2}\phi'\delta\phi',\\
        \label{eq: xi_pert} \left(\frac{e^{\chi} \delta\Xi'}{z^{2}}\right)'&= -\frac{1}{z^2}\left[\delta a'+\Delta_2\delta\chi' + \frac{2 \Delta_2\delta\chi}{z} - \phi' \Delta_2\delta\phi\right],\\
        \label{eq: f_pert} \left(\frac{\delta f}{z^{3}}\right)^{\prime}&=\frac{\delta\Xi}{z^{3}}+\frac{L^{2}}{2z^{4}}e^{-\chi}\left(\frac{dV(\phi)}{d\phi}\delta\phi-V(\phi)\delta\chi\right)-\left[\left(\frac{\delta\Xi}{2z^{2}}\right)^{\prime}+\frac{e^{-\chi}\delta a}{2z^{2}}+\frac{e^{-\chi}}{2z^{2}}(\Delta_{2}\delta\chi-2\delta\chi)\right]\\
        \delta\dot{a}'-\frac{\delta\dot{a}}{z}&=\frac{z^{2}}{2}\left(\frac{f\delta a^{\prime}}{z^{2}}\right)^{\prime}-\frac{z^{2}}{2}\left(\frac{1}{z^{2}}[\Delta_{2}\delta\Xi+2\delta\Xi]\right)^{\prime}-\frac{e^{-\chi}}{2}(\Delta_{2}^{2}\delta\chi+2\Delta_{2}\delta\chi),\\
        \delta\dot{\phi}'-\frac{\delta\dot{\phi}}{z}&=\frac{z^{2}}{2}\left(\frac{\phi'\delta f+f\delta\phi'}{z^{2}}\right)^{\prime}-\frac{\phi'\delta\Xi}{2}+\frac{e^{-\chi}\Delta_{2}\delta\phi}{2}-\frac{L^{2}}{2z^{2}}e^{-\chi}\left(\frac{d^2V(\phi)}{d\phi^2}\delta\phi-\frac{dV(\phi)}{d\phi}\delta\chi\right),\\
        e^{\chi}\delta\dot{\Xi}^{\prime}&=-\frac{2}{z}\Delta_{2}(\delta f+f\delta\chi)+\Delta_{2}(\delta f^{\prime}-\chi^{\prime}\delta f)-f(\delta a^{\prime}+\Delta_{2}\delta\chi^{\prime})+\delta\dot{a}+\Delta_{2}\delta\dot{\chi}+f\phi'\Delta_{2}\delta\phi+2\delta\Xi,\\
        \frac{2}{z}(\delta\dot{f}+f\delta\dot{\chi})&=e^{-\chi}\Delta_{2}\delta f-[f\delta\Xi]^{\prime}+f\chi^{\prime}\delta\Xi-2[\partial_{v}-f\partial_{z}]\delta\Xi+f\phi'\delta\dot{\phi}.
    \end{align}
\end{widetext}

Subsequently, we decompose $\delta\bm{\Phi}=(\delta\chi,\delta\Xi,\delta f,\delta a,\delta\phi)$ as $\tilde{\bm{\Phi}}(z)e^{-i\omega v}P_l(\cos\theta)$, where $\tilde{\bm{\Phi}}(z)=(\tilde{\chi}(z),\tilde{\Xi}(z),\tilde{f}(z),\tilde{a}(z),\tilde{\phi}(z))$ are the expansion coefficients and $P_l(\cos\theta)$ is the Legendre polynomial of order $l$. In this decomposition, the time derivatives and the two-dimensional Laplacian operators can be replaced by $-i\omega$ and $-l(l+1)$ respectively. Meanwhile, the $z$ coordinate is discretized with Chebyshev-Gauss-Lobatto grid points and the radial derivatives are replaced by the corresponding differentiation matrix. In this way, the quasi-normal frequencies $\omega$ with a specified angular quantum number $l$ can be obtained by solving a generalized eigenvalue problem:
\begin{equation}
    (\bm{A}+\omega\bm{B})\tilde{\bm{\Phi}}=0,
\end{equation}
where $\bm{A}$ and $\bm{B}$ depend on the background solution and $l$.

The equations we use to compute the quasi-normal modes are similar to the free evolution scheme described in Appendix \ref{appendix: general}. In this scheme, Eqs. (\ref{eq: xi_pert}) and (\ref{eq: f_pert}) are used to detect numerical errors and we do not need to add boundary conditions for energy and momentum conservation.
As standard, we take ingoing boundary conditions at the horizon, which means a regular solution in our coordinates. At the AdS boundary, Dirichlet or Neumann boundary conditions are imposed to be consistent with the background:
\begin{equation}
    \tilde{\chi}|_{z=0}=0,\quad\tilde{\Xi}|_{z=0}=0,\quad\tilde{a}|_{z=0}=0,\quad\tilde{\phi}'|_{z=0}=0.
\end{equation}

We use the function \texttt{scipy.linalg.eig} in Python to solve the generalized eigenvalue problem, obtaining the eigenvalues $\omega$ and the corresponding eigenfunction $\tilde{\bm{\Phi}}$. Substituting these numerical results into the redundant linearized equations (\ref{eq: xi_pert}) and (\ref{eq: f_pert}) yields the numerical errors. We use two criteria to test whether a computed eigenvalue is really a quasi-normal mode and not just a numerical artifact as in \cite{Jansen:2017oag}. First, the numerical errors should be less than $10^{-5}$. Second, we repeat the calculation with different grid sizes of the $z$ coordinate and test for convergence. We see that the lowest mode has been computed quite accurately, and as the mode number (both $n$ and $l$) increases, the accuracy of our result decreases. Fortunately, the higher modes are not important for the physics that we are interested in.

\subsection{Dynamical instability}

We have calculated quasi-normal modes for all equilibrium states in the phase diagrams Fig. \ref{fig: phase diagram} for angular quantum numbers ranging from $l=0$ to $l=5$. The tendency for the modes to vary with $l$ suggests that larger angular quantum numbers are less likely to give rise to unstable modes.
The states with unstable quasi-normal modes are marked in orange in Fig. \ref{fig: phase diagram}.
It can be seen that for the scalar source $\phi_1=1$ and $1.5$ there are no unstable states at all.
For the source $\phi_1=1.9$, unstable states are present, although only in a small part of the spinodal region. So there must exist a critical value of the source below which all black hole solutions are axially dynamically stable at the linear level. By dichotomizing, we find that this critical value is about $\phi_1=1.889$. On the other hand, the fact that some states located in the spinodal region are dynamically stable indicates a sharp contrast with the case of planar topology.
These states are free from dynamical instability, even though they suffer from thermodynamic instability, indicating the violation of the Gubser-Mitra conjecture in our case. Remarkably, the discrepancy between the dynamically unstable region and the spinodal region depends on the scalar source. Specifically, the larger the source, the more the dynamical and thermodynamic instabilities coincide.
Without loss of generality, we choose a supercritical value $\phi_1=2$ to show the inconsistency between dynamical and thermodynamic instabilities by displaying quasi-normal spectra and performing nonlinear dynamical evolution in Section \ref{sec: nonlinear}.

\begin{figure}
    \includegraphics[width=0.9\linewidth]{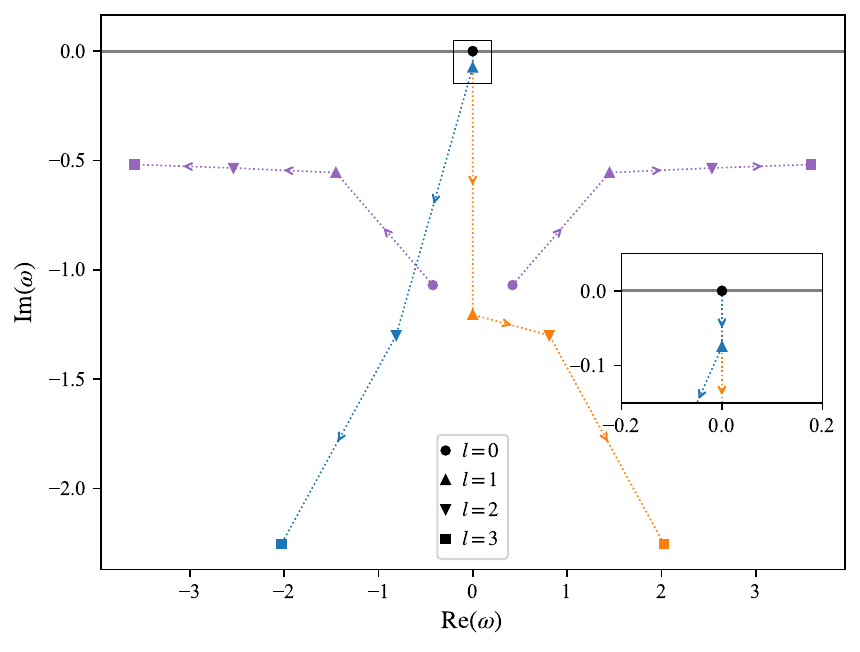}
    \includegraphics[width=0.9\linewidth]{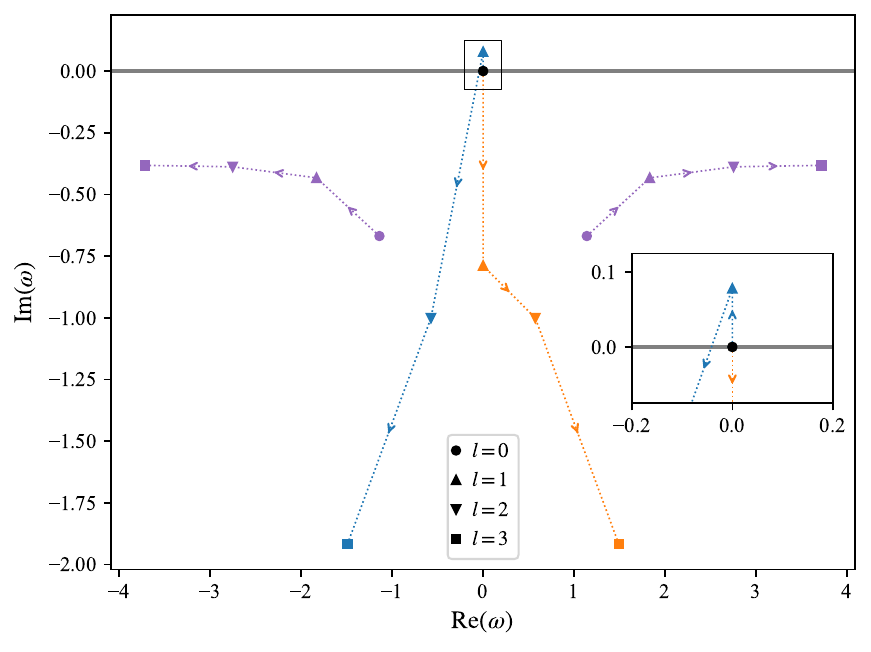}
    \caption{The quasi-normal modes of the equilibrium states represented by the green (upper panel) and red (lower panel) dots in Fig. \ref{fig: phase diagram} (d) with the variation of the angular quantum number $l$. The different colors and markers represent different quasi-normal modes and angular quantum numbers, respectively. The black dots indicate the common points of the two branches of hydrodynamic-like modes for $l=0$. The dotted lines connecting some of the markers represent the trajectories of the modes varying with the angular quantum number, and the arrows indicate the corresponding direction of migration.}
    \label{fig: QNM}
\end{figure}

For comparison, the green and red dots located in the spinodal region in Fig. \ref{fig: phase diagram} (d) are selected as initial data for the nonlinear dynamical evolution. The quasi-normal spectra for these configurations, with angular quantum numbers ranging from $l=0$ to $l=3$, are shown in Fig. \ref{fig: QNM}. From the figures, one can observe that {in both cases for small angular quantum numbers there are two branches of modes ending at the origin (represented by the black dots in the figures)}, similar to the hydrodynamic modes in the case of planar topology, where the modes go to the origin as the momentum $k$ goes to zero. Such purely imaginary modes become oscillating modes when $l$ exceeds a critical value, which depends on the energy density of the thermal phase.

On the other hand, for the configuration denoted by the green dot, all the modes lie on or below the real axis, indicating the dynamical stability at the linear level. The situation is distinct for the solution represented by the red dot, where the thermodynamic instability leads to the emergence of a type of axial instability. As shown in the lower panel of Fig. \ref{fig: QNM}, the imaginary part of the mode with the angular quantum number $l=1$ is positive in one of the hydrodynamic-like branches, indicating the dynamical instability under specific axial perturbations. Such a single unstable mode necessarily lacks oscillatory behavior due to the symmetry $\omega\rightarrow -\omega^*$ resulting from the real scalar field configuration.

\section{Nonlinear dynamical simulation}\label{sec: nonlinear}

\subsection{Numerical approach}

Based on the results of the linear analysis, we further investigate the nonlinear dynamics of the gravitational system to reveal the final fate of the dynamical instability. 
With the metric ansatz (\ref{eq: metric}), the gravitational dynamics is transformed into a time-dependent solution to a set of coupled partial differential equations. Due to the special nested structure, these equations can be solved sequentially with appropriate boundary conditions at the AdS boundary.

To simulate the gravitational system, we evolve the fields in a fixed computational domain $[0,z_0]$ in the $z$ direction, where $z_0$ must be slightly larger than the reciprocals of the apparent horizon radius at all times. Thus, $z_0$ is chosen a priori, and we need to reserve some space for the apparent horizon to evolve. We find that $z_0=1.3$ is a good choice for the initial states represented by the green and red dots in Fig. \ref{fig: phase diagram} (d). 
There is also previous work \cite{Balasubramanian:2013yqa} using the similar metric ansatz, which turned on an additional field and used the remaining gauge freedom to fix the apparent horizon at $z=1$.
Both horizon-fixed and horizon-unfixed schemes can work, with their own advantages and disadvantages. The horizon-unfixed method leads to a simpler formalism and we do not need to solve the elliptic equation representing the apparent horizon condition at each time step, which reduces the consumption of computational resources. However, the evolution of the field inside the apparent horizon introduces additional numerical errors. Fortunately, in the case of $\phi_1=2$, this error is acceptable, about $10^{-5}$. Additional numerical details, including the procedure for solving fully nonlinear equations, the boundary conditions, the procedure for detecting numerical errors, and the test for convergence are described in Appendices \ref{appendix: general} and \ref{appendix: nonlinear}.

We use the Chebyshev pseudospectral discretization in the $z$ direction as standard.
In the direction of $\theta$, we double the coordinate range from $[0,\pi]$ to $[-\pi,\pi]$. A similar operation is performed in the radial direction of the polar coordinates \cite{trefethen2000spectral}, from $r\in [0,1]$ to $r\in [-1,1]$, which avoids dealing with boundary conditions at the coordinate singularity $r=0$ and prevents the grid points from being too dense $r=0$.
In our case, the periodic boundary condition can also be employed therefore.
This allows us to use the Fourier pseudospectral discretization instead of the Chebyshev pseudospectral discretization, leading to several advantages. Firstly, it avoids having to deal with the boundary conditions at the north and south poles. Secondly, the grid points in the north and south pole regions are not as dense as in the Chebyshev spectrum, which reduces both the errors caused by the coordinate singularities there and the CFL instability at the same time step size. On the other hand, we utilize an even number of Fourier grid points, denoted as $M$, ranging from $-\pi+\pi/M$ to $\pi-\pi/M$ to avoid the coordinate singularity at the north and south poles. With this extension, the functions $\chi$, $f$, $A$, and $\phi$ exhibit even symmetry with respect to $\theta$, while $\xi$ displays odd symmetry.
Thus we can only evolve with data from $0$ to $\pi$.

\subsection{Apparent horizon condition}

While it is not necessary to solve the equation for the apparent horizon condition at the time of evolution, we do need to extract the horizon configuration at each instant from the time-dependent solutions, since we are interested in the shape of the black hole horizon as well as the entropy density.

There is a simple way to derive the apparent horizon condition \cite{Chesler:2013lia}. We define the congruence as $k_\mu(x)=\lambda(x)H(x)_{,\mu}$, where $x=(v,z,\theta,\varphi)$ and the surface $H(x)=0$ within a given time-slice denotes the apparent horizon. Considering the null condition $k^\mu k_\mu=0$, the time derivative $\partial_v H$ can be fixed in terms of the spatial derivatives of $H(x)$. Requiring the congruence to satisfy the affinely parameterized geodesic equation $k^\mu k_{\nu;\mu} = 0$ determines the time derivative of the rescaling function $\lambda(x)$ in terms of its spatial derivatives. With these time derivatives, the expansion can be calculated by $\Theta={k^\mu}_{;\mu}$, which must vanish on the apparent horizon $H(x)=0$, giving the apparent horizon condition.

In our case, we consider the apparent horizon $H=z-h(v,\theta)$, and the above steps give the following condition
\begin{widetext}
\begin{equation}
    (\partial_\theta+h_\theta\partial_{z})(\xi+h_\theta e^{-A-\chi})+\cot\theta(\xi+h_\theta e^{-A-\chi})=-\frac{1}{h}(f-h_\theta^{2}e^{-A-\chi}),
\end{equation}
\end{widetext}
where all fields (except $h$ and its $\theta$-derivative) should be understood as functions of $v$, $z=h(v,\theta)$, and $\theta$.
Using the pseudospectral discretization and barycentric interpolation, this differential equation can be cast into a linear equation, essentially a root-finding problem, which can be solved by the function \texttt{scipy.optimize.root} in Python.

After the apparent horizon configurations $z_h=h(\theta)$ are determined, it is convenient to define the entropy density from the Bekenstein-Hawking formula as follows:
\begin{equation}
    s(\theta)=\frac{2\pi}{h^2},
\end{equation}
and the total entropy can be obtained by integrating the entropy density over the angular direction:
\begin{equation}
    S=\int_0^{2\pi}\int_0^\pi s(\theta)\sin\theta d\theta d\varphi.
\end{equation}

\subsection{Simulation results}

\begin{figure}
    \includegraphics[width=0.9\linewidth]{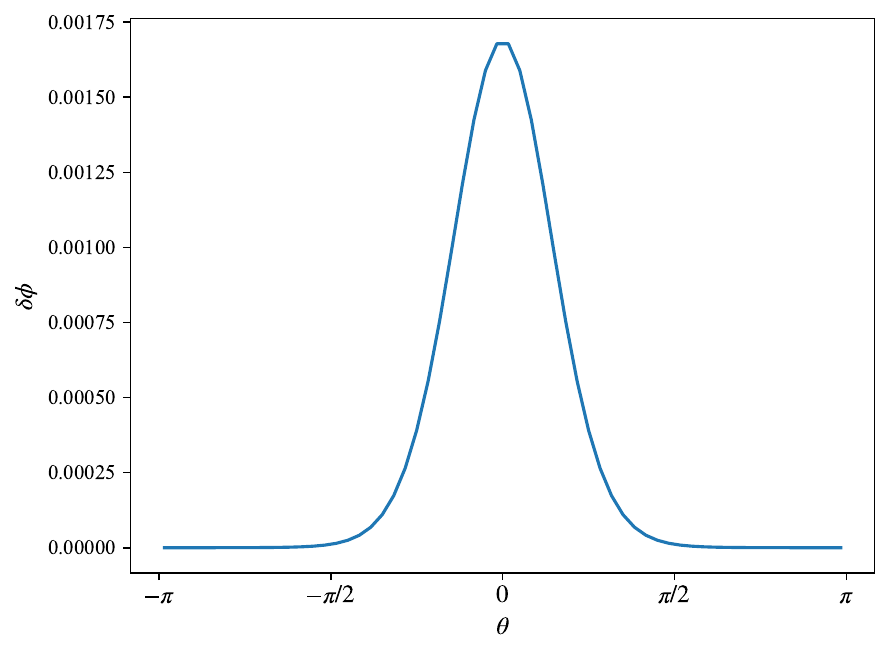}
    \caption{The angular configuration of the scalar perturbation with amplitude $\phi_0=0.001$ at the inner boundary $z=z_0$.}
    \label{fig: perturbation configuration}
\end{figure}

We consider the initial states represented by the green and red dots in Fig. \ref{fig: phase diagram} (d) and introduce a $\theta$-dependent perturbation to the scalar field, the form of which is chosen as a mixture of all modes without loss of generality:
\begin{equation}
    \delta\phi(z,\theta)=\phi_0 z^2 \exp\left(-10\sin^2\frac{\theta}{2}\right),
\end{equation}
where $\phi_0$ indicates the amplitude of the perturbation. For the state represented by the green dot, the amplitude of the perturbation ranges from $\phi_0=0.1$ to $\phi_0=1$, while for the red dot, the range is $\phi_0=0.001$ to $\phi_0=0.01$. The perturbation configuration with amplitude $\phi_0=0.001$ at the inner boundary $z=z_0$ is shown in Fig. \ref{fig: perturbation configuration}.

We use 50 Chebyshev-Gauss-Lobatto grid points in the $z$ direction and 60 Fourier grid points in the $\theta$ direction (30 points in the physical range $[0,\pi]$) in our numerical simulations. We also run the simulations with other numbers of grid points and perform a convergence test, shown in Appendix \ref{appendix: nonlinear}. The time step size is taken as $\Delta v=0.0005$ and the fourth-order Runge-Kutta method is used to evolve the fields in time. The initial data are evolved until $v=200$, which is long enough for the dynamical transition to occur.

\begin{figure}
    \includegraphics[width=\linewidth]{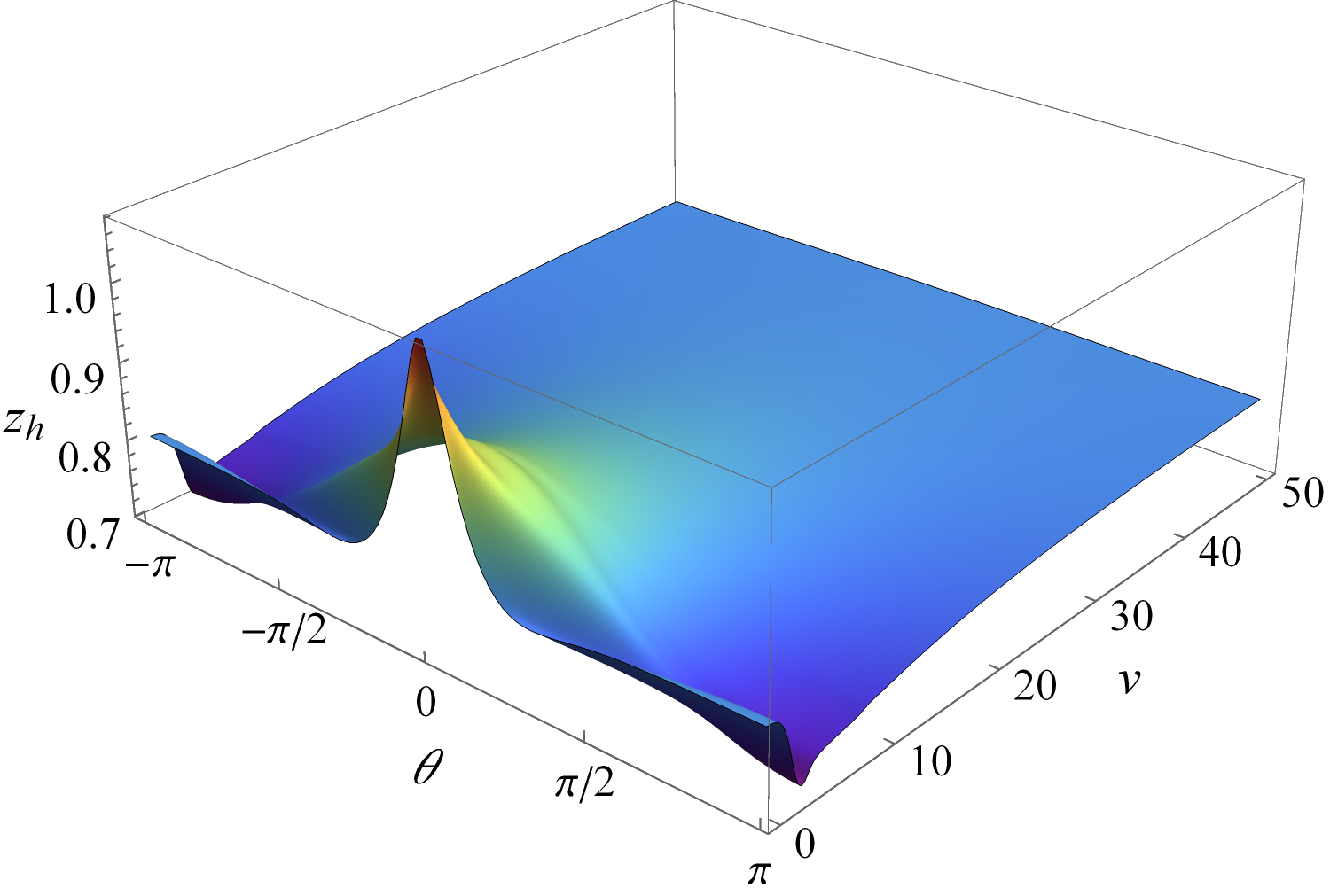}
    \caption{The temporal evolution of the apparent horizon configuration for the initial state denoted by the green dot.}
    \label{fig: damping}
\end{figure}

The fully nonlinear dynamical simulations demonstrate that the gravitational system in the state denoted by the green dot resists such an axial perturbation, indicating the dynamical stability, consistent with the linear analysis.
We take the perturbation amplitude $\phi_0=1$ as an example, and the temporal evolution of the apparent horizon configuration is shown in Fig. \ref{fig: damping}. It can be seen that the axial perturbation damps with time, leaving a spherically symmetric black hole.

\begin{figure}
    \includegraphics[width=0.9\linewidth]{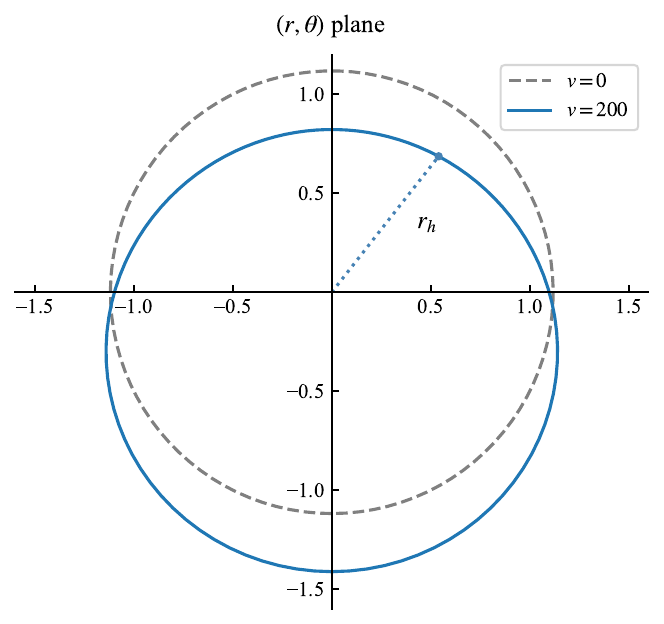}
    \includegraphics[width=0.9\linewidth]{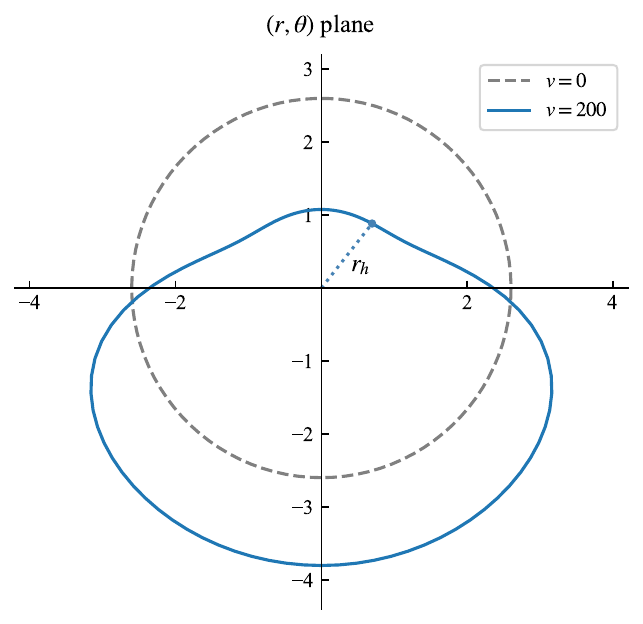}
    \caption{The $\varphi=\mathrm{const}$ plane of the apparent horizon for the initial and final states of the spontaneous deformation for the scalar source $\phi_1=2$ (upper panel) and $\phi_1=4$ (lower panel).}
    \label{fig: horizon_polar}
\end{figure}

For the initial state denoted by the red dot, we take the amplitude of the perturbation to be $\phi_1=0.001$.
In stark contrast to the former, here the unstable mode with angular quantum number $l=1$ is excited under the axial perturbation, leading to drastic changes in the gravitational configuration.
The $\varphi=\mathrm{const}$ plane of the apparent horizon for the initial and final states is depicted in the upper panel of Fig. \ref{fig: horizon_polar}. One can observe that the apparent horizon radius $r_h$ decreases in the north pole region and increases in the south pole region, eventually leading to the formation of a black hole with only axial symmetry.
For larger sources, such as $\phi_1=4$, the deformation will be more obvious, as shown in the lower panel of Fig. \ref{fig: horizon_polar}. However, we find that the numerical errors in the dynamical evolution increase for larger sources, and are about $10^{-3}$ for $\phi_1=4$. Therefore, we show it only for schematic purposes and leave the case of larger sources for future work.

\begin{figure}
    \includegraphics[width=0.9\linewidth]{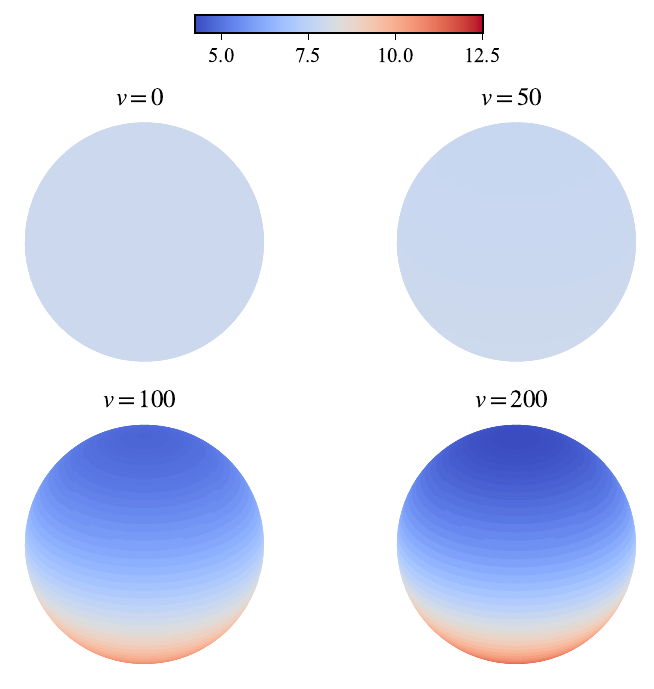}
    \caption{The angular dependence of the entropy density at different times for the initial state represented by the red dot. The color spectrum indicates the value of the entropy density.}
    \label{fig: configuration_3d}
\end{figure}

\begin{figure}
    \includegraphics[width=0.9\linewidth]{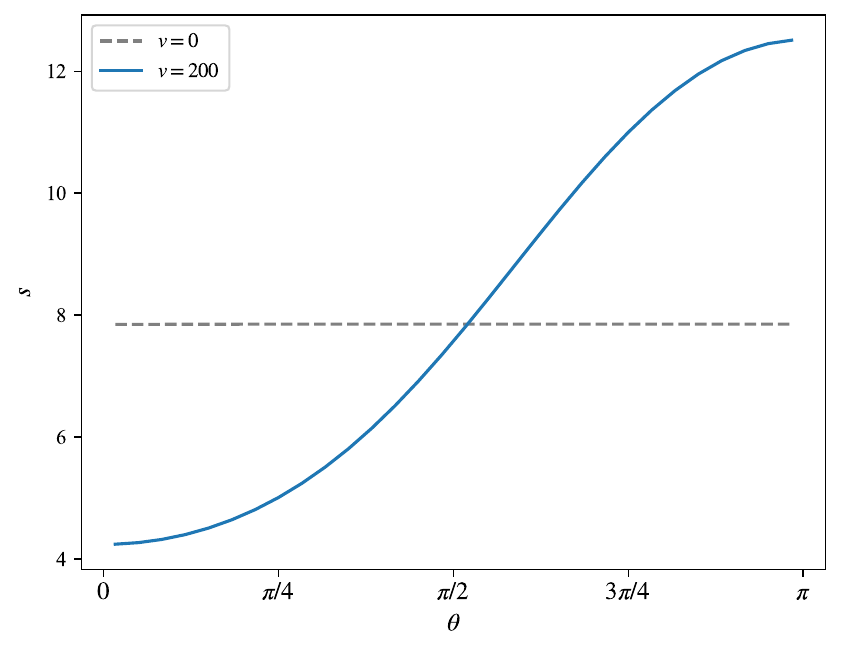}
    \caption{The angular dependence of the final entropy density configuration for the initial state represented by the red dot.}
    \label{fig: final entropy density}
\end{figure}

\begin{figure}
    \includegraphics[width=0.9\linewidth]{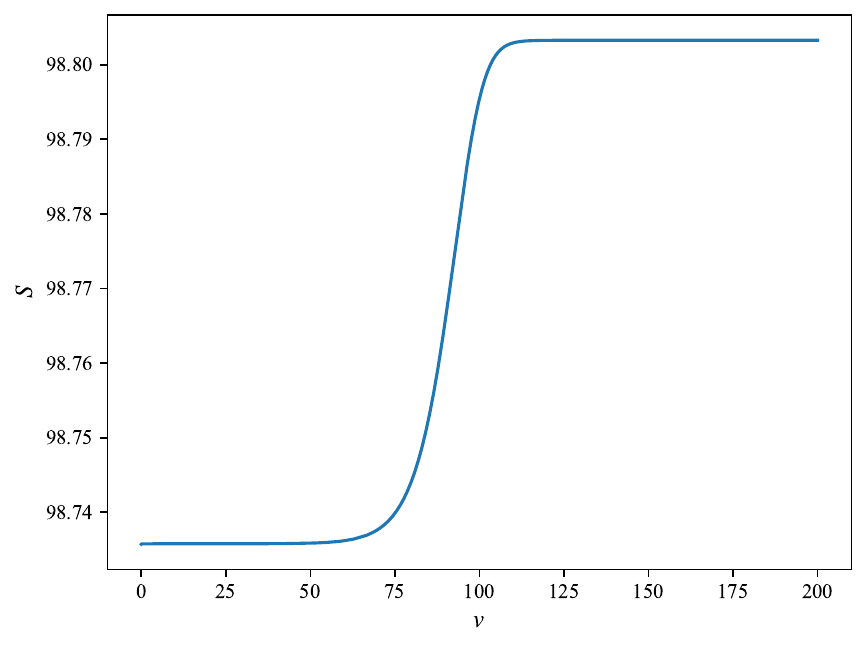}
    \caption{The temporal evolution of the total entropy for the initial state denoted by the red dot.}
    \label{fig: total entropy}
\end{figure}

As the horizon configuration evolves, the distribution of the entropy density shows a similar behavior, as shown in Fig. \ref{fig: configuration_3d}, which displays the angular dependence of the entropy density configurations at different times during the dynamical transition, where the color spectrum indicates the value of the entropy density.
The final entropy density configuration is shown in Fig. \ref{fig: final entropy density}, clearly demonstrating a strong $\theta$-dependence in the final state. Although the entropy density exhibits distinct dynamical behavior in different angular regions during the intermediate dynamical process, the total entropy always increases monotonically with time, as shown in \ref{fig: total entropy}, consistent with the second law of black hole mechanics. On the other hand, we have checked that the scalar source remains isotropic throughout the time evolution, thus it can be concluded that the spherical symmetry of the gravitational system is broken spontaneously, resulting in a dynamical deformation process.

\section{Conclusion}\label{sec: conclusion}

In this paper, we have revealed the real-time dynamics of a spontaneous dynamical transition from a spherically symmetric black hole to an axisymmetric black hole by working with a certain gravitational model in the presence of an isotropic external scalar source. Such an initial black hole with spherical symmetry lies deep in a region of thermodynamic instability in the phase diagram. Different from the case of planar topology, where the entire spinodal region is linearly dynamically unstable, in our case only the central part of the spinodal region satisfies the Gubser-Mitra conjecture, indicating a kind of axial dynamical instability. We find that the energy range of this dynamically unstable region depends on the scalar source. For sources too small, there are no dynamically unstable states. For sources above the critical value, the larger the source, the larger the parameter range of dynamical instability in the spinodal region. In the intermediate process of the dynamical evolution of this instability, there is a drastic change in the apparent horizon $r_h$, manifested by an inward contraction of the north pole region and an outward expansion of the south pole region. Eventually a black hole with an event horizon with angular dependence forms as the final fate of this dynamical instability.
However, the scalar source remains isotropic throughout the dynamical evolution, indicating that this type of dynamical deformation spontaneously breaks the spherical symmetry.

For future research, a natural direction would involve exploring the real-time dynamics of the spontaneous deformation of rotating black holes.
It will be interesting to study whether the black hole spin inhibits or promotes this dynamical transition.
The thermodynamic relationships involving the axisymmetric static solutions also deserve to be revealed, visualizing the competition between the spherically symmetric and axisymmetric solutions.
Another question deserving further attention is the dynamical behaviors of the modes during the transition, especially for larger values of the scalar source, which may give rise to more unstable modes.

\begin{acknowledgments}
    We would like to thank Zhoujian Cao, Xiao-Kai He, Bin Wang, and Cheng-Yong Zhang for their helpful discussions. YT would like to thank Yong-Ming Huang for helping obtain the EOM in the general Bondi-Sachs-like coordinates. This work is partly supported by the National Key Research and Development Program of China (Grant No.2021YFC2203001). This work is supported in part by the National Natural Science Foundation of China under Grants No. 11975235, No. 12035016, No. 12075026, and No. 12275350.
\end{acknowledgments}

\appendix

\section{General formalism of numerical relativity under the Bondi-Sachs-like metric}\label{appendix: general}

In this Appendix, we denote time derivatives with respect to $v$ using dots or subscripts $v$, radial derivatives with respect to $z$ using primes, and angular derivatives with respect to $\theta$ using subscripts $\theta$.

The most general ingoing form of metric under the Bondi-Sachs-like gauge in $1+d$ dimensions can be expressed as
\begin{widetext}
\begin{equation}\label{eq: general_metric}
    ds^{2}=\frac{L^{2}}{z^{2}}(-fe^{-\chi}dv^{2}-2e^{-\chi}dvdz+h_{ij}[dx^{i}-\xi^{i}dv][dx^{j}-\xi^{j}dv]),
\end{equation}
\end{widetext}
where $f$, $\chi$, $h_{ij}$, and $\xi^i$ are functions of $(v,z,x^i)$ and $i$, $j$ range from $1$ to $d-1$. In the asymptotically AdS or dS cases, the constant $L$ can be conveniently set to the AdS or dS radius, while for the asymptotically flat case, $L$ can be arbitrarily chosen. Additionally, the determinant of $h_{ij}$ is constrained everywhere to be that of the standard sphere, plane, hyperbolic space, or even more complicated topologies in higher dimensions, and so $z$ is proportional to the inverse of the local areal radius.

Some general discussion of this metric will be provided in detail elsewhere. Here we only list the corresponding Einstein equations for the planar case (with $\det(h_{ij})=1$):
\begin{widetext}
\begin{equation}\label{eq: general_chi_constraint}
    S_1(\Psi)=\frac{d-1}{z}\chi^{\prime}-\frac{1}{4}h^{ik}h^{jl}h_{li}^{\prime}h_{kj}^{\prime},
\end{equation}
\begin{equation}\label{eq: general_xi_constraint}
    S_2(\Psi)=-\frac{d-1}{z}\chi_{,i}-\Theta_{i,j}^{j\prime}-\frac{1}{2}h_{kj}^{\prime}h_{,i}^{kj}-z^{d-1}\left(\frac{\Theta_{ij}\xi^{j\prime}}{z^{d-1}}\right)^{\prime},
\end{equation}
\begin{equation}\label{eq: general_f_constraint}
    \begin{aligned}
        S_3(\Psi) &= -z^{d-1}\left(\frac{2f}{z^{d}}\right)^{\prime}+\xi_{,i}^{i\prime}-z^{d-1}\left(\frac{2\xi_{,i}^{i}}{z^{d-1}}\right)^{\prime}+\frac{1}{4}h^{kn}\left(2h_{nj,i}-h_{ij,n}\right)h_{,k}^{ij}e^{-\chi}\\
        &\quad -\left(h_{,i}^{ij}\chi_{,j}+h^{ij}\chi_{,ij}-\frac{1}{2}h^{ij}\chi_{,j}\chi_{,i}\right)e^{-\chi}+h^{ij}_{,ij}e^{-\chi}+\frac{1}{2}\xi^{i\prime}\Theta_{ij}\xi^{j\prime},
    \end{aligned}
\end{equation}
\begin{equation}\label{eq: general_h_evolution}
    \begin{aligned}
        S_4(\Psi)	&\simeq	\dot{h}_{j(n}h_{m)}^{j\prime}+\frac{1}{2}\dot{h}_{mn}^{\prime}+z^{d-1}\left(\frac{\dot{h}_{mn}}{2z^{d-1}}\right)^{\prime}-\frac{1}{2}f h_{in}^{\prime}h_{m}^{i\prime}-z^{d-1}\left(\frac{fh_{mn}^{\prime}}{2z^{d-1}}\right)^{\prime}+h_{i(n}^{\prime}\partial_{m)}\xi^{i}+z^{d-1}\left(\frac{h_{i(m}\partial_{n)}\xi^{i}}{z^{d-1}}\right)^{\prime}\\
        &\quad	+\frac{1}{2}(h_{mn}^{\prime}\xi^{k})_{,k}+z^{d-1}\left(\frac{\xi^{k}h_{mn,k}}{2z^{d-1}}\right)^{\prime}-\frac{1}{4}(2h_{(nj,i}-h_{ij,(n})h_{,m)}^{ij}e^{-\chi}+\left[ h^{kl}\left(\partial_{(m}h_{ln)}-\frac{1}{2}\partial_{l}h_{mn}\right)e^{-\chi}\right]_{,k}\\
        &\quad	+\frac{1}{2}\chi_{,m}\chi_{,n} e^{-\chi}+(\chi_{,m} e^{-\chi})_{,n}-\frac{1}{2}h_{mj}h_{ni}\xi^{j\prime}\xi^{i\prime}e^{\chi},
    \end{aligned}
\end{equation}
\begin{equation}\label{eq: general_xi_evolution}
    \begin{aligned}
        S_5(\Psi) &= -\frac{1}{2}(\Theta_{kj}\xi^{j\prime})_{,v}+\frac{1}{2}(d_{n}\Theta_{k}^{j}+h^{ij}h_{lk}\xi_{,i}^{l}-\xi_{,k}^{j}-\xi^{j}\Theta_{ki}\xi^{i\prime})_{,j}\\
        &\quad +\frac{f_{,k}^{\prime}}{2}-\frac{d-1}{2z}(f_{,k}+f\chi_{,k})+\frac{1}{4}h_{,k}^{ij}d_{n}h_{ij}-\frac{1}{2}\Theta_{l,k}^{i}\xi_{,i}^{l}-\frac{1}{2}\xi_{,k}^{i}\Theta_{ij}\xi^{j\prime},
    \end{aligned}
\end{equation}
\begin{equation}\label{eq: general_f_evolution}
    \begin{aligned}
        S_6(\Psi)&=\frac{1}{2}(-[f\xi^{i}]^{\prime}+\xi^{i}\xi^{j\prime}\Theta_{jk}\xi^{k}-\xi^{k}d_{n}\Theta_{k}^{i}+\Theta^{ij}f_{,j}-2d_{n}\xi^{i}+\xi^{k}\xi^{i}_{,k}-\xi^{m}\Theta_{mk}\Theta^{ij}\xi^{k}_{,j})_{,i}+\frac{1}{4}\dot{h}^{ij}d_{n}h_{ij}\\
        &\quad-\frac{1}{2}\dot{\Theta}_{k}^{j}\xi_{,j}^{k}+\frac{1}{2}\xi^{i}(\Theta_{ij}\xi^{j\prime})_{,v}-\frac{d-1}{2z}(\dot{f}+f\dot{\chi}),
    \end{aligned}
\end{equation}
\end{widetext}
where $h^{ij}$ is the matrix inverse of $h_{ij}$, $h_j^{i\prime}:=h^{ik}h_{kj}^\prime$, $\Theta_{ij}:=e^\chi h_{ij}$, $\Theta^{ij}$ is the matrix inverse of $\Theta_{ij}$, $\partial_\mu\Theta_j^{i}:=\Theta^{ik}\partial_\mu\Theta_{kj}$, $d_{n}:=\partial_{v}-f\partial_{z}+\xi^{i}\partial_{i}$, $\Psi$ denotes the matter content of the theory, $S_k(\Psi)$ $(k=1,2,\dots,6)$ represent the terms depending on the matter fields, and $\simeq$ means equality up to the trace part of a tensor equation. For other topologies, the corresponding equations can be obtained by appropriate coordinate transformations. In particular, for the spherical case in the four dimensions we consider in this paper, the EOM can be achieved by the transformation $x=-\cos\theta$ in one of the two transverse directions. These equations have a very nice nested structure.\footnote{The matter part may ruin the nested structure, but fortunately the scalar field we consider in this paper does not.} Due to the fact that they are not fully independent, two possible numerical schemes can be adopted in the dynamical evolution.

We take the asymptotically AdS case as an example \cite{Balasubramanian:2013yqa}. Initially, the data for the metric component $h_{ij}$, the material fields $\Psi$, and some integration constants on the time slice $v_0$ should be given. These integration constants are the coefficients of the cubic term in the near-boundary expansions of the fields $\xi^i$ and $f$, denoted as $\xi_3^i$ and $f_3$ respectively. This data enables us to treat Eq. (\ref{eq: general_chi_constraint}) as a first-order linear ordinary differential equation for the field $\chi$. By choosing the boundary condition $\chi|_{z=0}=0$, we can easily solve for $\chi$. Once the field values of $h_{ij}$, $\Psi$, and $\chi$ are determined, the quantities ${\xi^i}'$ and $\xi^i$ can be consecutively obtained by solving Eq. (\ref{eq: general_xi_constraint}), and the corresponding boundary conditions are ${\xi^i}'''|_{z=0}=6\xi_3^i$ and the gauge choice $\xi^i|_{z=0}=0$, respectively. Next, we solve Eq. (\ref{eq: general_f_constraint}) to obtain the field $f$, with the boundary condition $f'''|_{z=0}=6f_3$. Subsequently, the field $\dot{h}_{ij}$ are determined by solving Eq. (\ref{eq: general_h_evolution}) with the boundary condition $\dot{h}_{ij}'|_{z=0}=0$. Then, the matter field equations should be solved with appropriate boundary conditions. Using these field values, we integrate $\dot{h}_{ij}$ and $\dot{\Psi}$ over time to push $h_{ij}$ and $\Psi$ to the next time slice $v_0+dv$. Subsequently, the field value of $\chi$ on the time slice $v_0+dv$ can be obtained in the same way described above. And then there are two options about the fields $\xi^i$ and $f$, leading to two different evolution schemes:

\begin{itemize}
    \item the constrained scheme:

    The time derivatives of $\xi_3^i$ and $f_3$ can be determined through the near-boundary expansion of Eqs. (\ref{eq: general_xi_evolution}) and (\ref{eq: general_f_evolution}). Then, these integration constants can be updated to the time slice $v_0+dv$ by integrating over time. Consequently, we can solve for $\xi^i$ and $f$ on the time slice $v_0+dv$ in the same manner described above. Finally, all fields on the time slice $v_0+dv$ are determined and the evolution of a time step $dv$ is completed. This iterative procedure continues until the simulation is complete. Two redundant equations (\ref{eq: general_xi_evolution}) and (\ref{eq: general_f_evolution}) are included in the process to identify numerical errors.

    \item the free evolution scheme:
    
    The time derivatives of $\xi^i$ and $f$ can be obtained from Eqs. (\ref{eq: general_xi_evolution}) and (\ref{eq: general_f_evolution}) ($\dot{\chi}$ should be determined from the time derivative of Eq. (\ref{eq: general_chi_constraint}) in advance). Then, we can obtain $\xi^i$ and $f$ on the time slice $v_0+dv$ by integrating $\dot{\xi}^i$ and $\dot{f}$ over time. Finally, all fields on the time slice $v_0+dv$ are determined and the evolution of a time step $dv$ is completed. In the subsequent evolution, we no longer need to solve Eqs. (\ref{eq: general_xi_constraint}) and (\ref{eq: general_f_constraint}). This iterative procedure continues until the simulation is complete. Two redundant equations (\ref{eq: general_xi_constraint}) and (\ref{eq: general_f_constraint}) are included in the process to identify numerical errors.
\end{itemize}
The constrained scheme with the constraints imposed at some other boundaries instead of the AdS conformal boundary, as well as the free evolution scheme, does not sensitively depend on the asymptotics of the spacetime and so can be also applied to asymptotically flat and dS cases (though certain technical subtleties may still arise).

\section{Numerical procedure for dynamical evolution}\label{appendix: nonlinear}

In this paper, we focus on the axisymmetric and non-rotating case, and the ingoing Bondi-Sachs-like metric (\ref{eq: general_metric}) in four-dimensional asymptotically AdS spacetime\footnote{See also, e.g. \cite{Winicour:2005eoq} for a review of dynamical evolution under the Bondi-Sachs-like gauge in four-dimensional asymptotically flat spacetime.} degenerates to
\begin{widetext}
\begin{equation}
    ds^2=\frac{L^2}{z^2}(-[fe^{-\chi}-e^{A}\xi^2]dv^2-2e^{-\chi}dvdz-2\xi e^Advd\theta+e^Ad\theta^2+e^{-A}\sin^2\theta d\varphi^2),
\end{equation}
\end{widetext}
where $L$ is set to the unit, and all metric components and the scalar field depend on $v$, $z$, and $\theta$. To aid our calculations, we introduce the following auxiliary variables
\begin{widetext}
\begin{equation}
    \begin{aligned}
        P&=\frac{e^{A+\chi}}{4}\xi^{\prime2}+\frac{\xi_{\theta}}{z}-\frac{\xi_{\theta}^{\prime}}{2}-\frac{e^{-A-\chi}}{4}(2A_\theta\chi_\theta+\chi_{\theta}^2-\phi_\theta^2)+\frac{(e^{-A-\chi})_{\theta\theta}}{2},\\
        Q&=\frac{\xi'}{2}-\frac{\xi}{z}+\frac{e^{-A-\chi}\chi_\theta}{2}.
    \end{aligned}
\end{equation}
\end{widetext}
As a result, the EOM are simplified to
\begin{widetext}
\begin{align}
    \label{eq: chi_constraint} \chi'&=\frac{z}{4}(A'^2+\phi'^2),\\
    \label{eq: xi_constraint} \left(\frac{e^{A+\chi}}{z^2}\xi'\right)'&=-\frac{1}{z^{2}}\left[(A+\chi)^{\prime}_{\theta}+\frac{2\chi_{\theta}}{z}-(A^{\prime}A_{\theta}+\phi'\phi_{\theta})+2\cot\theta A'\right],\\
    \label{eq: f_constraint} \left(\frac{f}{z^{3}}\right)^{\prime}&=\frac{\xi_{\theta}}{z^{3}}+\frac{L^{2}}{2z^{4}}e^{-\chi}V(\phi)+\frac{1}{z^{2}}\left[P-\cot\theta\left(Q-\frac{\xi}{z}+\frac{3e^{-A-\chi}A_\theta}{2}\right)-e^{-A-\chi}\right],\\
    \label{eq: A_evolution} \dot{A}^{\prime}-\frac{\dot{A}}{z}&=\frac{z^{2}}{2}\left(\frac{fA^{\prime}-\xi A_{\theta}}{z^{2}}\right)^{\prime}+\frac{(e^{-A-\chi}A_{\theta}-\xi A^{\prime})_{\theta}}{2}+P+\cot\theta\left(Q-\frac{\xi A'}{2}\right),\\
    \label{eq: phi_evolution} \dot{\phi}^{\prime}-\frac{\dot{\phi}}{z}&=\frac{z^{2}}{2}\left(\frac{f\phi'-\xi\phi_{\theta}}{z^{2}}\right)^{\prime}+\frac{(e^{-A-\chi}\phi_{\theta}-\xi\phi')_{\theta}}{2}-\frac{L^{2}}{2z^{2}}e^{-\chi}\frac{dV(\phi)}{d\phi}+\cot\theta\left(\frac{e^{-A-\chi}\phi_\theta}{2}-\frac{\xi \phi'}{2}\right),\\
    (e^{A+\chi}\xi^{\prime})_{v}&=-\frac{2}{z}(f_{\theta}+f\chi_{\theta})-(2\xi e^{A+\chi}\xi^{\prime}-f^{\prime}+f[A^{\prime}+\chi^{\prime}]-[\dot{A}+\dot{\chi}])_{\theta}+\xi(e^{A+\chi}\xi^{\prime}+A_\theta+\chi_\theta)_{\theta}-A_{\theta}d_{n}A-\phi_{\theta}d_n\phi \nonumber \\
    &\quad-\cot\theta[\xi(e^{A+\chi}\xi'-3A_\theta+\chi_\theta)+2f A'-2\dot{A}]+2\xi, \label{eq: xi_evolution} \\
    \frac{2}{z}(\dot{f}+f\dot{\chi})&=(e^{-A-\chi}f_{\theta}-[f\xi]^{\prime}+\xi^{2}e^{A+\chi}\xi^{\prime}-\xi[d_{n}A+d_{n}\chi]-2d_{n}\xi)_{\theta}-\xi_\theta(\dot{A}+\dot{\chi})-\dot{A}d_{n}A-\dot{\phi}d_n\phi+\xi(e^{A+\chi}\xi^{\prime})_{v} \nonumber \\
    &\quad+\cot\theta(\xi^2[e^{A+\chi}\xi'-(A_\theta+\chi_\theta)]+\xi[f(A'+\chi')-2\dot{\chi}-2\xi_\theta-f']+f\xi'-2\dot{\xi}+e^{-A-\chi}f_\theta), \label{eq: f_evolution}
\end{align}
\end{widetext}
where $d_{n}=\partial_{v}-f\partial_{z}+\xi\partial_{\theta}$. There is a systematic and efficient integration strategy to solve these equations, which benefits from their nested structure. We use the constrained scheme described in Appendix \ref{appendix: general} in this paper. More specifically, the boundary conditions arise from the near-boundary asymptotic behaviors of the field solutions:
\begin{align}
    \label{eq: chi_asym} \chi&=\frac{\phi_1^2}{8}z^2+O(z^3),\\
    \label{eq: xi_asym} \xi&=\xi_3 z^3+O(z^4),\\
    \label{eq: f_asym} f&=1+\left(\frac{\phi_1^2}{8}+1\right)z^2+f_3 z^3+O(z^4),\\
    \label{eq: A_asym} A&=O(z^3),\\
    \label{eq: phi_asym} \phi&=\phi_1 z+\phi_2 z^2+O(z^3).
\end{align}

For Eq. (\ref{eq: chi_constraint}), the boundary condition $\chi|_{z=0}=0$ is just a gauge choice.
For Eq. (\ref{eq: xi_constraint}), the asymptotic behavior of the field $\xi$ (\ref{eq: xi_asym}) yields two boundary conditions with the gauge choice $A|_{z=0}=0$:
\begin{equation}
    \left.\left(\frac{e^{A+\chi}}{z^2}\xi'\right)\right|_{z=0}=3\xi_3,\quad\xi|_{z=0}=0.
\end{equation}
The asymptotic behavior of the field $f$ (\ref{eq: f_asym}) determines a single integration constant in Eq. (\ref{eq: f_constraint}). By imposing the field redefinition $f=1+z^2\tilde{f}$, the corresponding boundary condition becomes $\tilde{f}'|_{z=0}=f_3$. Moreover, the time derivatives of the integration constants $\xi_3$ and $f_3$ are determined through the near-boundary expansion of Eqs. (\ref{eq: xi_evolution}) and (\ref{eq: f_evolution}):
\begin{equation}
    \begin{aligned}
        \partial_v\xi_3&=\partial_\theta\left(\frac{ f_3}{3}-A_3-\frac{2\phi_1\phi_2}{9}\right)-2\cot\theta A_3,\\
        \partial_v f_3&=\frac{3\partial_\theta\xi_3}{2}+\frac{\phi_1\partial_v\phi_2}{6}+\cot\theta\frac{3\xi_3}{2}.
    \end{aligned}
\end{equation}
Hence, the values of $\partial_v\xi_3$ and $\partial_v f_3$ can be calculated after obtaining $\dot{A}$ and $\dot{\phi}$ from Eqs. (\ref{eq: A_evolution}) and (\ref{eq: phi_evolution}).
Then $A$, $\phi$, $\xi_3$, and $f_3$ can be updated to the next time slice by integrating over time.

\begin{figure*}
    \subfigure[]{\includegraphics[width=0.45\linewidth]{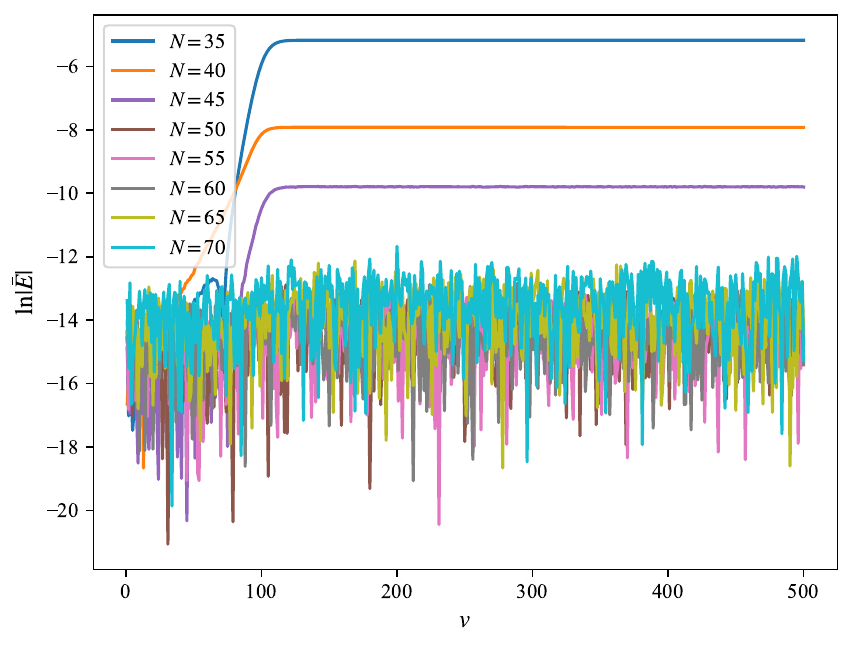}}
    \subfigure[]{\includegraphics[width=0.45\linewidth]{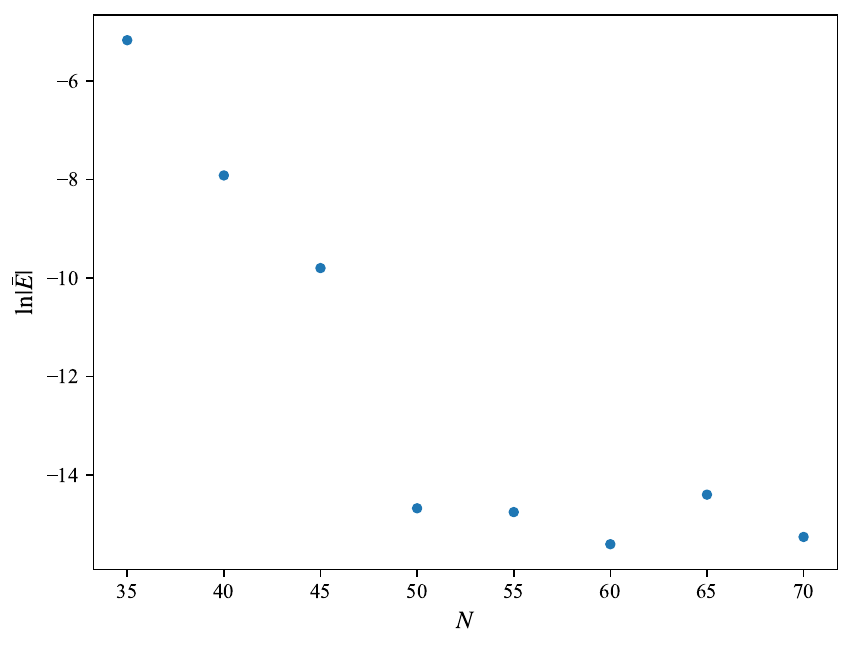}}
    \subfigure[]{\includegraphics[width=0.45\linewidth]{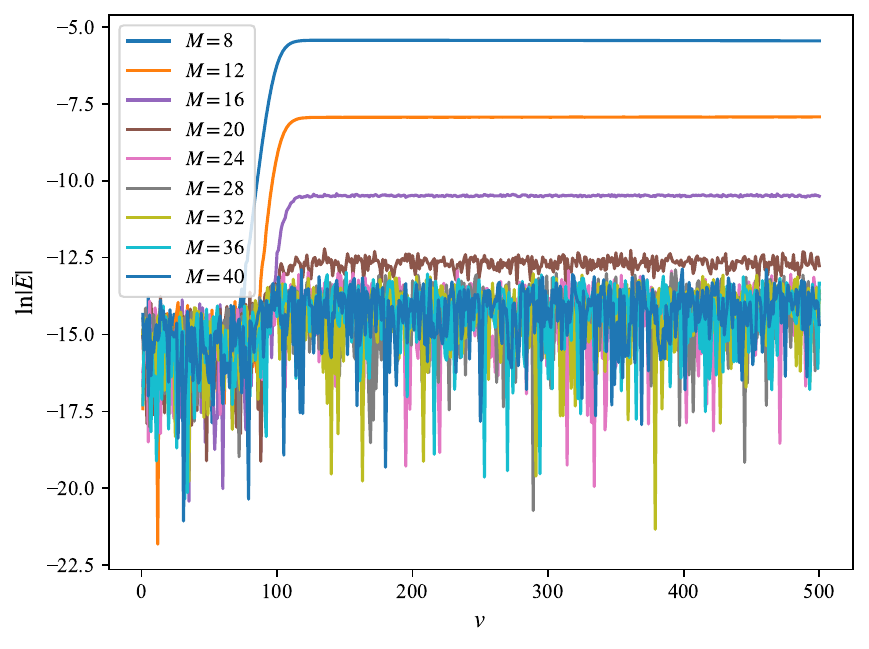}}
    \subfigure[]{\includegraphics[width=0.45\linewidth]{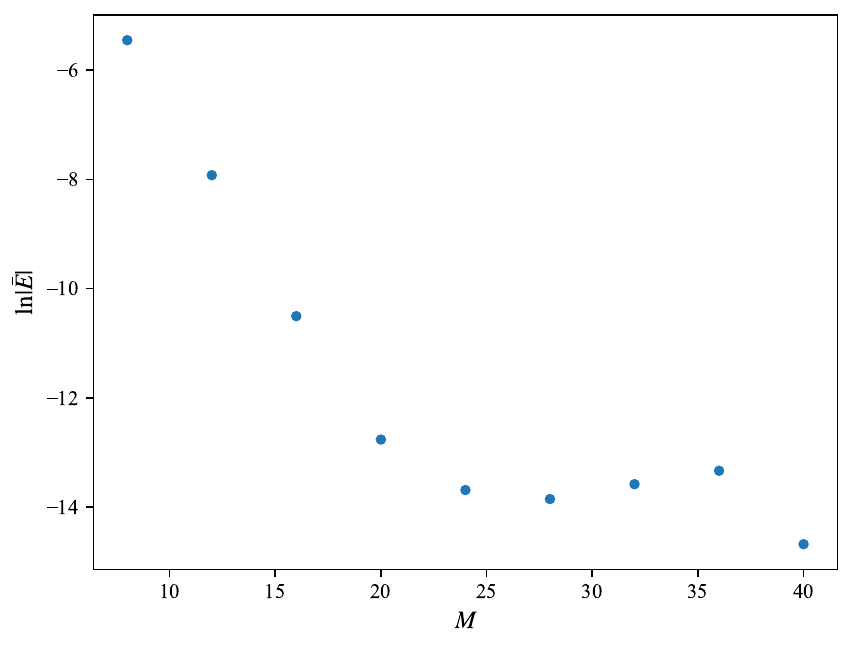}}
    \caption{Convergence test of our numerical code during the spontaneous deformation in Fig. \ref{fig: horizon_polar}. Left panel: the temporal evolution of the value of $\ln|\bar{E}|$ with different numbers of grid points $N$ in the radial direction (at this moment we take $M=40$) (a) and different numbers of grid points $M$ in the angular direction (at this moment we take $N=50$) (c). Right panel: the value of $\ln|\bar{E}|$ at $v=500$ varies with the number of grid points $N$ in the radial direction (b) and $M$ in the angular direction (d).}
    \label{fig: convergence test}
\end{figure*}

In such a constrained scheme, two redundant equations (\ref{eq: xi_evolution}) and (\ref{eq: f_evolution}) can be used to identify numerical errors. Since their asymptotic behaviors in the near-boundary region are already used to determine the boundary conditions, we can use the values of Eq.(\ref{eq: xi_evolution}) at the apparent horizon to detect numerical errors, which are averaged along the $\theta$ direction and denoted as $\bar{E}$. We take the case of spontaneous deformation in Fig. \ref{fig: horizon_polar} as an example to test the convergence of our numerical code, which is implemented in Python (Numpy and Scipy). We vary the number of grid points in the $z$ direction or in the $\theta$ direction and keep the other parameters unchanged. The results are shown in Fig. \ref{fig: convergence test}, as the temporal evolution of the value of $\ln|\bar{E}|$ and its value at $v=500$ varying with the number of grid points. It can be seen that the accuracy of our numerical code improves exponentially as the number of grid points increases, which is exactly what is expected from the spectral method \cite{Chen:2022tfy,trefethen2000spectral}.

\bibliography{ref.bib}

\end{document}